\documentclass[twocolumn,showpacs,preprintnumbers,amssymb]{revtex4}
\usepackage{graphicx}

\begin{document}

\title{A supernova constraint on bulk majorons}

\author{Steen Hannestad~\footnote[1]{email: steen@nordita.dk}, 
Petteri Ker{\"a}nen~\footnote[2]{keranen@nordita.dk}, 
Francesco Sannino~\footnote[3]{francesco.sannino@nbi.dk}}

\affiliation{NORDITA, Blegdamsvej 17, DK-2100 Copenhagen, Denmark}

\date{\today}

\begin{abstract}
In models with large extra dimensions all gauge singlet fields can
in principle propagate in the extra dimensional space. We have
investigated possible constraints on majoron models of neutrino
masses in which the majorons propagate in extra dimensions. It is
found that astrophysical constraints from supernovae are many
orders of magnitude stronger than previous accelerator bounds.
Our findings suggest that
unnatural types of the ``see-saw" mechanism for neutrino masses are
unlikely to occur in nature, even in the presence of extra dimensions.
\end{abstract}


\maketitle

\section{Introduction}

Over the past few years there has been an enormous interest in models
with extra dimensions beyond the 3+1 dimensions of the standard model.
Such extra dimensions can be large if the gauge fields of the standard
model are constrained to a 3+1 dimensional brane in the higher dimensional
bulk space. This idea is particularly interesting because it may be
connected with a low quantum gravity scale, perhaps only slightly
above the scale of electroweak symmetry breaking.
In Refs.~\cite{add98,Antoniadis,add99}
such a model was proposed in which the standard model brane
is embedded in a bulk space of dimension $\delta$
which is toroidally compactified.
Because gravity is allowed to propagate in the full bulk space it
will look weaker to an observer confined to the brane. By Gauss' law
the effective four-dimensional Planck scale, $M_P$, can be related to the
true 4+$\delta$ dimensional energy scale of gravity, $M$, by
\begin{equation}
M_P^2 = (2 \pi R)^\delta M^{\delta+2},
\end{equation}
where $R$ is the common radius of all the extra dimensions.
If $R$ is sufficiently large $M$ can be as small as the electroweak
scale. Effectively gravity is weak because the field lines leak into
the extra dimensions.

One of the most interesting features of these models is the
presence of a tower of new modes for particles propagating in the
bulk. The momentum of particles propagating in extra dimensions is
discretised because the extra dimensions are compact. To an
observer on the brane this effectively looks like a tower of new
states (Kaluza-Klein) for each bulk particle, with a 4D mass
related to the extra dimensional momentum. The energy spacing
between these states is then in general of order $R^{-1}$ which
can be very small.

For the graviton this property has been used to put tight constraints
on the possible radii of the extra dimensions. However, any particle
which is a singlet in the standard model could in principle propagate
in the bulk. Examples of this are sterile neutrinos and axions.

Some authors, for example, have investigated the possibility that
right handed neutrinos might propagate in the extra dimensions and
that their resultant suppressed couplings to the usual left handed
neutrinos could account for the low scale of neutrino
masses~\cite{DDG,ADDM,Mohapatra:1999zd,Mohapatra:1999af,Dvali:1999cn,Barbieri:2000mg,Mohapatra:2000wn,Ioannisian:mu}. 
Other authors~\cite{FP,MN,DK,C,BDN,Abazajian:2000hw,Ioannisian:1999cw} 
have
studied the constraints on such models due to experiment and
observation. An alternate approach~\cite{MPP}, corresponding to the
conventional see-saw mechanism, has also been investigated by some
authors. In this approach a Higgs singlet, which carries no
standard model gauge quantum numbers, is allowed to propagate in
the extra dimensions and give mass to right handed neutrinos
which, for simplicity, are assumed to live on the brane.

Recently in Ref.~\cite{MNSS} a model was studied in which the
lepton number is spontaneously broken and the associated Goldstone
boson (the ``Majoron") is present  (see also~\cite{MRS}). This
model was used to compute the decay rate for the intermediate
vector boson $Z$ to two neutrinos and the Majoron (denoted by $J$)
or one of its ``Kaluza-Klein" excitations~\cite{MNSS}. Using the
accurately known value of the $Z$ width~\cite{rpp} bounds on the
dimensionless coupling of the two neutrinos to the Majoron were
obtained allowing for any number of extra dimensions and any
intrinsic mass scale (see also \cite{Carone:2002ss}). 
The related neutrinoless double beta decay
process $n+n\rightarrow p+p+e^{-}+e^{-}+(J)$ has already been
treated in a model of the present type~\cite{MPP}. A detailed
discussion of supernova constraints in the 3 $+$ 1 dimensional
theory has very recently been given in~\cite{TPV}.

In this paper we use this model~\cite{MNSS} to compute processes
of astrophysical interest. We first see that when compact extra
dimensions are present new processes become relevant and can
heavily affect supernovae dynamics. We then show that supernovae
constraints on the dimensionless couplings are by many order of
magnitude more stringent than the accelerators bounds. Our
findings seem to discourage unnatural see-saw models of neutrino
masses still allowed by accelerator bounds.

We summarize in section II the Majoron model extended to the
extra-dimensional brane-world . In section III we briefly review
the accelerator constraints. The Supernovae constraints  are
presented in section IV. In the concluding section V we also
discuss the possibility of the extra-dimensional Majorons to be
source of dark matter.


\section{A Majoron Model in Extra Dimensions}

In the original Majoron model of Chikashige, Mohapatra and Peccei~\cite{CMP}
the lepton number associated with massive Majorana
neutrinos is spontaneously broken. Here, the notations of
Ref.~\cite{SV} and Ref.~\cite{MNSS} for the $3+1$ and the extra
dimensional case will be followed, respectively. In addition to the usual Higgs
doublet
\begin{equation} \phi = \left(
\begin{array}{c}
\phi ^{+} \\ \phi ^{0}
\end{array}
\right) \qquad l=0 \label{usualhiggs}
\end{equation}
 which has lepton number $l$ equal to zero, the model contains
an electrically neutral complex singlet field
\begin{equation}
\Phi \qquad l=-2 \ . \label{singlethiggs}
\end{equation}
It is required that the Higgs potential constructed from $\Phi$
and $\phi$ conserves lepton number. The vacuum expectation values are:
\begin{eqnarray}
\langle\Phi \rangle &=&\langle \Phi^{\ast} \rangle \ ,  \nonumber \\
\langle\phi ^{0}\rangle &=&\langle{\phi^{0}}^{\dagger}
\rangle=\lambda \approx 2^{-\frac{1}{4}}G_{F}^{-\frac{1}{2}}\ ,
\label{vacvalues}
\end{eqnarray}
where $G_F$ is the Fermi constant and $\langle\Phi\rangle$
(whose non-zero value violates lepton number) sets a
new scale in the theory.

The three light physical two component neutrinos
$\nu_{1},\nu_2,\nu_3$ acquire Majorana masses $m_1,m_2,m_3$ which
are of order $\epsilon^2 {\cal M}_{H}$ with
\begin{equation}
\epsilon = {\cal O} \left( \frac{\cal D}{{\cal M}_H}\right) \ .
\label{eps}
\end{equation}
According to the a standard ``see-saw mechanism"~\cite{ss}
$\displaystyle{{\cal D}/{\lambda}}$ represents the ``Dirac-type"
coupling constants for the bare light neutrinos while the $3\times
3$ matrix $\displaystyle{{{\cal M}_H}/{\langle\Phi\rangle}}$
represents the Majorana type coupling constants for the bare heavy
(or ``right handed") neutrinos. Assuming the heavy scale ${\cal
M}_H$ to be substantially larger than the energy in play in the
following we focus our attention only on the light neutrinos. The
coupling of the Majoron $J$, identified as $J={\rm Im}\,\Phi$, to
the physical neutrino fields $\nu_1,\nu_2,\nu_3$ in $3+1$
dimensions is:
\begin{equation}
{\cal L}_{J}=i\frac{J}{2}\sum_{a,b=1}^3 \nu^T_{a}\sigma_2 g_{ab}
\nu_b + {\rm h.c.} \ . \label{lj1}
\end{equation}
It turns out~\cite{SV} that the coupling constants have the
expansion:
\begin{equation}
g_{ab}=-\frac{1}{\langle\Phi \rangle}m_a \delta_{ab} + {\cal
O}\left(\epsilon^4 {\cal M}_{H} \right) \ , \label{gexp}
\end{equation}
where the leading term is diagonal in generation space. It is
convenient to express this leading term using four component
ordinary Dirac spinors
\begin{equation}\psi_{a} =\left(
\begin{array}{c}
\nu _{a } \\ 0\end{array} \right) \ , \label{4comp}
\end{equation}
in a $\gamma_5$ diagonal representation of the Dirac matrices the
relevant Lagrangian term reads:
\begin{equation}
{\cal L}_{J}=i\frac{J}{2\langle\Phi \rangle}\sum_{a=1}^3 m_a
\left(\psi^T_a C^{-1}\frac{1+\gamma_5}{2}\psi_a +
\bar{\psi}_a\frac{1-\gamma_5}{2}C \bar{\psi}^T_a\right) \ .
\label{lj2}
\end{equation}
Here $C$ is the charge conjugation matrix of the Dirac theory.

The generalization of the present model to the case where the
field $\Phi$ propagates in $\delta$ extra dimensions has been
carried out in detail in Ref.~\cite{MNSS}. These extra dimensions,
denoted as $y_{i}$ with $i=1,\ldots \delta$, are assumed to be
toroidally compactified with a radius $R_i$.
For simplicity we assume that all the radii $R_i$ are equal to the
same value $R$. $\Phi\left(x,y\right)$ continues to carry the
``engineering dimension" one as it would in $3+1$ dimensional
space-time and via a Fourier expansion with  respect to the
compactified coordinates we have:
\begin{eqnarray}
\Phi\left(x,y\right)={\rm Norm}\sum_{n_1,\ldots,n_{\delta}}
\Phi_{n_1,\ldots, n_{\delta}}\left(x\right)e^{\frac{i}{R}\left(n_1
y_1+ \cdots \right)} \ , \label{normalization}
\end{eqnarray}
where $\displaystyle{{\rm Norm}=\left[2\pi M
R\right]^{-\frac{\delta}{2}}}$ and $M$ represents the intrinsic
scale of the new theory.

Each general Kaluza-Klein field receives a mass squared increment
\begin{equation}
\Delta m^2_{n_1,n_2,\ldots}=\frac{1}{R^2}\left(n_1^2+n_2^2+\cdots
n_{\delta}^2 \right) \ . \label{increment}
\end{equation} The zero-mass Majoron
$J_{0,0,\ldots,0}(x)$ corresponds to the previously studied $3+1$
massless Majoron. The fields ${\rm Re}\, \Phi_{n_1,n_2,\ldots}(x)$
are expected to receive a substantial increment from the pure
Higgs sector of the theory~\cite{MNSS} and will be neglected in
the following.

The intrinsic scale $M$ and the compactification radius $R$ can
be related to each other when assuming in the ``brane" model the
graviton to propagate in the full $(3+\delta)+1$ dimensional
space-time. Then the ordinary form of Newtons' gravitation law is
only an approximation valid at distances much greater than $R$.
The Newtonian gravitational constant (inverse square of the Planck
mass $M_P$) is obtained~\cite{add98,{Antoniadis},{add99}} as a
phenomenological parameter from
\begin{equation}
\left(\frac{M_P}{M}\right)^2=\left(2\pi M
R\right)^{\delta}=\frac{1}{\left(\rm Norm\right)^2} \ .
\label{Planck}
\end{equation}
Considering $M_P$ as an experimental input (and approximating
$R_1=R_2=\cdots =R_{\delta}$), shows via (\ref{Planck}) that $M$
is the only free parameter introduced to describe the extra
dimensional aspect of the present simple theory when $\delta$ is
fixed.

We expect ${\cal M}_H/\langle\Phi \rangle$ to be very roughly of
the order unity and $\langle\Phi \rangle$ of the order of $M$.
Finally the Yukawa interactions of the Majoron and its
Kaluza-Klein excitations with the light neutrinos are described by
(c.f. (\ref{lj2})):
\begin{eqnarray}
{\cal L}_{J}&=&\sum_{a=1}^3 \sum_{n_1,\ldots,n_{\delta}} i
g_{aa;n_1,\ldots,n_{\delta}} J_{n_1,\ldots,n_{\delta}} \times
\nonumber
\\&&\times\left(\psi^T_a C^{-1}\frac{1+\gamma_5}{2}\psi_a +
\bar{\psi}_a\frac{1-\gamma_5}{2}C \bar{\psi}^T_a\right) \ ,
\label{lj3}
\end{eqnarray}
to leading order in the neutrino masses, $m_a$ with
\begin{equation}
g_{aa;n_1,\ldots,n_{\delta}}\equiv g_{aa}=\frac{m_a}{2\langle\Phi
\rangle} \frac{M}{M_P} \ .
\end{equation}
The vacuum expectation value $\langle \Phi \rangle$ (unless
unnatural fine tuning is considered) is very roughly of the order
of $M$, leading $g_{aa}$ to be naturally of the order of
$\displaystyle{{m_a}/{M_P}}\approx 10^{-28}$ regardless of the
number of extra dimensions~\cite{MNSS}. Clearly the exact value of
this universal coupling crucially depends on the unknown dynamics
behind the spontaneous breaking of the lepton number~\cite{MNSS}.

\section{Review of Constraints from Accelerators} In~\cite{MNSS}
it has been shown that the accurately known value of the Z width
can provide information about the coupling of two neutrinos to the
Majoron. Both the $3+1$ dimensional case and the case in which one
adopts a ``brane" world picture with the Majoron free to
experience the extra dimensions have been studied. Bounds on the
dimensionless coupling constants were obtained, allowing for any
number of extra dimensions and any intrinsic mass scale.
If the uncertainty of the $Z$'s invisible width $1.7 \times
10^{-3}$~GeV is roughly taken as an indication of the maximum
allowed value for the total width into a Majoron and two neutrinos
the following bounds on $\mid g_{aa} \mid$ were obtained
in~\cite{MNSS} for $M_S = 10^4~GeV$:
$\mid g_{aa} \mid <3.4 \times 10^{-12}$, $\mid g_{aa} \mid < 2.3
\times10^{-10}$, $\mid g_{aa} \mid < 1.5\times10^{-8}$ for
$\delta=2,3,4$, respectively. These are much stronger bounds than
the one obtained for the model in $3+1$ space-time dimensions
which is $\mid g_{aa} \mid < 0.11$.

If a technically natural see-saw model is adopted, the predicted
coupling constants are far below these upper bounds. In addition,
for this natural model, the effect of extra dimensions is to
decrease the predicted partial Z width, the increase due to many
Kaluza-Klein excitations being compensated by the decrease of
their common coupling constant.

We shall see in the following that constraints from supernovae are
much more stringent than the ones above.


\section{Constraints from supernovae}

{\it Supernova cooling ---}
The proto-neutron stars created by core-collapse supernovae are born
with core temperatures of 30-50 MeV. The main cooling mechanism for these
stars is thermal surface emission of neutrinos on a timescale of a few
seconds. The total energy emitted is the order a few $10^{53}$ erg, with
roughly equal amounts in all flavours of neutrinos and antineutrinos.
This emission has been observed from SN1987A by Kamiokande~\cite{Hirata:ad},
IMB~\cite{Bratton:ww} and Baksan~\cite{baksan},
all observing $\bar\nu_e$ events. The results from SN1987A are compatible
with theoretical supernova models and therefore
put a constraint on any non-standard cooling mechanism carrying
away too much energy, and since majorons from the KK-tower will be produced
in the supernova and carry away energy 1987A data can be used to
constrain $g_{aa}$. This has been done several times in the literature
for the 3+1 dimensional majoron models~\cite{Aharonov:ee,Aharonov:ik,Choi:1987sd,Grifols:1988fg,Berezhiani:gf,Berezhiani:za,Choi:1989hi,Chang:1993yp,Pilaftsis:1993af,Berezhiani:1994jc,Kachelriess:2000qc,Tomas:2001dh}.
For these models, only a relatively
small band in parameter space is excluded for the following reason:
For large $g_{aa}$ the majorons are tightly coupled inside the neutron
star, and only escape via surface emission, like neutrinos. Therefore
they cannot carry away most of the energy and supernovae do not yield
significant constraints. On the other hand, for small $g_{aa}$ the
majorons do escape freely, but they are not produced in significant numbers.
The end result is that a band of roughly $3 \times 10^{-7}
< g_{aa} < 2 \times 10^{-5}$ is excluded~\cite{Kachelriess:2000qc}.

For the $3+1+\delta$ models the situation is different because each
KK-mode is very weakly coupled. Therefore we are always in the limit
where majorons escape freely once they are produced, and do not need
to worry about possible surface emission.

A fairly robust constraint on such ``bulk emission'' is the one
proposed by Raffelt~\cite{Raffelt:wa}, that the emissivity of the
neutron star medium should be
\begin{equation}
\epsilon \lesssim 10^{19} \,\, {\rm erg} \, {\rm g}^{-1} \, {\rm s}^{-1}.
\end{equation}

In the 3+1 dimensional case where majorons are strictly massless
the neutrino pair annihilation $\nu \bar\nu \to J$ is not allowed
kinematically and the most important processes are $\nu \bar\nu
\to JJ$ and $\nu \to \bar\nu J$. However, in the present case,
most of the emitted majorons have mass of order $T$, and the
simple pair-annihilation is by far the most important process. The
squared and spin-averaged matrix element for this process (per
single Kaluza-Klein excitation) is
\begin{equation}
|A|^2 = \frac{1}{2} g_{aa}^2 p_1 \cdot p_2.
\end{equation}
The emissivity per volume of the medium is then
\begin{eqnarray}
Q(m_J) & = & \int {d\tilde{p}_1}{d\tilde{p}_2}{d\tilde{p}_J}(2\pi)^4
f_\nu(p_1)f_\nu(p_2)
\nonumber \\
&& \,\, \times |A|^2 \delta^4(p_1+p_2-p_J) (E_1+E_2),
\end{eqnarray}
where $d\tilde p = d^3 {\bf p}/(2\pi)^3 2E$ and $f_\nu$ are 
the thermal distributions of the incoming neutrinos, 
$f_\nu \equiv e^{-p_\nu/T}$.
Doing the integral gives
\begin{equation}
Q(m_J) = \frac{1}{128 \pi^3} g_{aa}^2 T^5 x^4 K_2(x) \,\, , \,\,\,
x \equiv m_J/T \ ,
\end{equation}
where $K_2(x)$ is a Bessel function. However, we also need to sum
over the KK-tower in order to obtain the total emissivity
\begin{eqnarray}
Q & = &\frac{\pi^{\delta/2}}{\Gamma(\delta/2) (2\pi)^{\delta}}
\frac{M_{P}^2 T^{5+\delta}}{M^{2+\delta}}g_{aa}^2  \int_0^\infty dx x^3 K_2(x) \\
& = & \delta \pi^{(\delta-6)/2} 2^{\delta-5}
\frac{M_{P}^2 T^{5+\delta}}{(2\pi)^\delta M^{2+\delta}}
g_{aa}^2 \Gamma(3+\delta/2)
\end{eqnarray}
Using the rough equality $Q = \epsilon \langle \rho \rangle$,
$\langle \rho \rangle \simeq 4 \times 10^{14} \,\, {\rm g} \, {\rm cm}^{-3}$,
this can
be translated into a bound on $g_{aa}$ and $M$ which is
\begin{equation}
g_{aa} \lesssim X M_{\rm TeV}^{1+\delta/2} T_{30}^{-(5+\delta)/2},
\label{Eq:cooling}
\end{equation}
with the following values for $X$
\begin{equation}
X = \cases{1.0 \times 10^{-21} & for $\delta=2$ \cr 2.1 \times
10^{-17} & for $\delta=4$ \cr 4.6 \times 10^{-13} & for
$\delta=6$}
\end{equation}
In the above, $T_{30} = T/(30 \,\, {\rm MeV})$. In all figures we use
$T_{30} = 1$.
The bounds are shown in Fig.~1 as a function of $M$. However,
from considerations of graviton emission, strong bounds on $M$ already
exist~\cite{Cullen:1999hc,Barger:1999jf,hanhart,hanhart2,Hannestad:2001jv,Hannestad:2001xi}. The most recent, and most restrictive,
limits are those from Ref.~\cite{Hannestad:2001xi}.
These are shown as the thicker lines in the left part of the
figure. Values of $M$ in this range are already excluded from
graviton emission arguments~\cite{Hannestad:2001xi}.

{\it Old neutron star excess heating ---} When considering
graviton emission a much stronger bound on $M$ can be obtained by
considering the decays of KK-gravitons produced in supernovae.
Gravitons have a significant branching ratio into photons and the
decays therefore produce gamma rays. All cosmological supernovae
have therefore by the present produced a diffuse cosmic gamma
background. Comparing this with observations of the diffuse gamma
background by the EGRET instrument~\cite{egret} yields a bound on
$M$ significantly stronger than found from the supernova cooling
argument~\cite{Hannestad:2001jv}. More interestingly, the tightest
constraint comes from observations of old, isolated neutron stars.

Most of the KK-modes emitted from the proto-neutron star have
masses of order $3T$ and therefore also relatively low
velocities. This again
means that a large fraction (roughly one half) of the KK-modes
have velocities lower than the escape velocity, and that neutron
stars retain a halo of KK-modes with a typical radius of 2-3
$R_{NS}$. When these gravitons decay they heat the neutron star
and can lead to excess surface emission from old neutron stars.
This argument applies equally to KK-gravitons and majorons. For
gravitons it was used in Ref.~\cite{Hannestad:2001xi} to put an
extremely strong limit on $M$, $M > 1600$ TeV $(\delta=2)$, 70 TeV
$(\delta=3)$.

The KK-majorons only decay into neutrinos and not photons.
However, the typical energy of the emitted neutrinos is 50-100
MeV, and the neutron star is not transparent to neutrinos of such
high energy. Therefore the neutrinos hitting the neutron star
surface will heat it just like photons do. The surface luminosity
of the neutron star at late times should therefore reach a
constant level corresponding to the energy deposited by decay
neutrinos. This luminosity is given by
\begin{equation}
L_{NS} = f_{J} F_{J} E_{TOT} \langle \Gamma_J \rangle \frac{R_{NS}^2}{R_{halo}^2},
\end{equation}
where $f_{J}$ is the fraction of the total supernova  energy
emitted in majorons, $F_{J}$ is the fraction of the emitted
majorons remaining bound to the neutron star, taken to be $1/2$,
$E_{TOT}$ is the total SN energy, taken to be $3 \times 10^{53}$
erg, $\langle \Gamma \rangle$ is the average majoron decay rate,
and $R_{halo}$ is the typical radius of the majoron halo, taken to
be 2 $R_{NS}$. The decay rate for non-relativistic majorons is
given by
\begin{equation}
\Gamma_J = \frac{1}{64 \pi} g_{aa}^2 m_J.
\label{Eq:decay}
\end{equation}
$f_{J}$ is a function of $M$ and $g_{aa}$, and can be found from the
above cooling bound. The cooling bound, to a good approximation, corresponds
to half the total energy being emitted in majorons ($f_{J} \simeq 1/2$).
For gravitons the strongest bound comes from the neutron star PSR
J0953+0755~\cite{0953,psc96,ll99},
which is the oldest neutron star for which the thermal
surface temperature has been measured. Its total surface luminosity
is estimated to be $L \sim 10^{-5} L_\odot$~\cite{ll99}.
For majorons this neutron
star also yields a strong constraint on $g_{aa}$ which is roughly
\begin{equation}
g_{aa} \lesssim Y M_{\rm TeV}^{(2+\delta)/4} T_{30}^{-(6+\delta)/4},
\label{Eq:heating}
\end{equation}
with the following values for $Y$
\begin{equation}
Y = \cases{3.3 \times 10^{-22} & for $\delta=2$ \cr 4.8 \times
10^{-20} & for $\delta=4$ \cr 7.0 \times 10^{-18} & for
$\delta=6$}
\end{equation}
In all cases this bound is significantly stronger than the cooling
bound, just as it is for gravitons. The bounds are summarized in
Fig.~\ref{fig2}.  However, there is a limit to the applicability
of the bound. The age of the neutron star PSR J0953+0755 is
estimated to be $\tau = 1.7 \times 10^7$ yr~\cite{ll99}. If the
majorons decay faster than this, they will have vanished already
and cannot act as a heating source. From Eq.~(\ref{Eq:decay}) one
finds a lifetime of
\begin{equation}
\tau = 1.3 \times 10^{-21} \, g_{aa}^{-2} \, \left\langle
\frac{100\, {\rm MeV}}{m_J} \right\rangle \,\, {\rm s},
\end{equation}
giving a rough upper limit on $g_{aa}$ of $1.5 \times 10^{-18}$,
above which the limit of Eq.~(\ref{Eq:heating}) does not apply.

\begin{figure}[h]
\begin{center}
\includegraphics[width=7truecm]{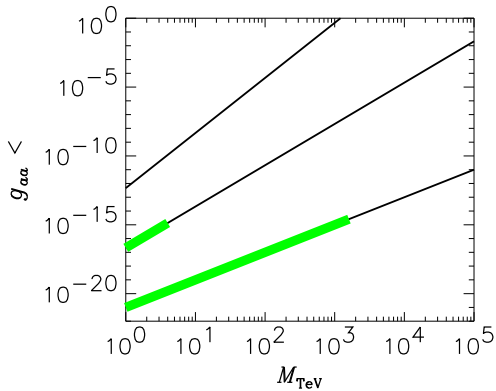}
\end{center}
\caption{Upper bounds on $g_{aa}$ from the supernova cooling
bound, Eq.~(\ref{Eq:cooling}), for various $\delta$ and $M$. The
bottom curve is for $\delta=2$, the middle for $\delta=4$, and the
top for $\delta=6$. The thick lines at low $M$ correspond to the
excluded region of $M$ from graviton effects.}
\label{fig1}
\end{figure}


\begin{figure}[h]
\begin{center}
\includegraphics[width=7truecm]{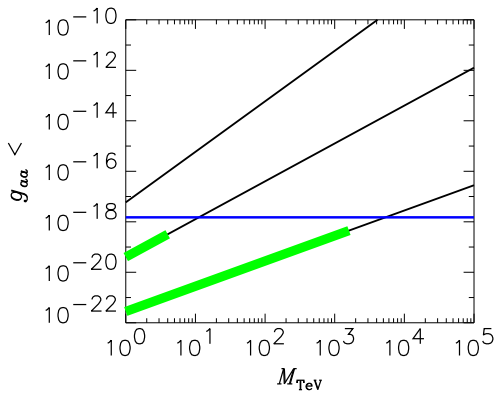}
\end{center}
\caption{Upper bounds on $g_{aa}$ from the neutron star heating
bound, Eq.~(\ref{Eq:heating}), for various $\delta$ and $M$. The
bottom curve is for $\delta=2$, the middle for $\delta=4$, and the
top for $\delta=6$. The thick lines at low $M$ correspond to the
excluded region of $M$ from graviton effects. The horizontal line
corresponds to the upper limit of applicability of the neutron
star heating limit.} \label{fig2}
\end{figure}

Another possible problem is that majorons could be reabsorbed when
they pass through the neutron star on a timescale much shorter than
the age of the neutron star \cite{hr02}.
The most relevant process for reabsorption is inverse bremsstrahlung,
$J NN \to NN$ which is induced via the electroweak interactions
\cite{CMP}. Since the majoron is a pseudo-scalar one can estimate
the rate for this process in the same way as for axions
\cite{Raffelt:wa}. The result is that reabsorption only happens on
a timescale much longer than the lifetime of the neutron star PSR
J0953+0755.


\section{Discussion}

\subsection{Majorons as dark matter?}

By the same neutrino pair annihilation process as in supernovae,
majorons are also produced in the early universe. This means
that in principle one might obtain non-trivial bounds on $g_{aa}$
from considering cosmological production of majorons. For gravitons
such considerations lead to extremely stringent bounds on $M$ and
the maximum temperature, $T_{RH}$, of the radiation dominated epoch
after inflation~\cite{hs99,Hannestad:2001nq,Fairbairn:2001ct,Fairbairn:2001ab}.
If the fundamental scale $M$ is close to 1 TeV then the reheating
temperature needs to be extremely low, typically in the MeV regime.
However, there is a lower limit to how low $T_{RH}$ can be without
disturbing big bang nucleosynthesis. Detailed calculations have shown
that this limit is roughly $T_{RH} \gtrsim 0.7$ MeV~\cite{Kawasaki:1999na,Kawasaki:2000en}.
The number density of majorons in the universe can be found by
solving the integrated Boltzmann equation
\begin{equation}
\dot{n}_J = - 3 H n_J + \frac{g_{aa}^2}{128 \pi^3} m_J^3 T K_1(m_J/T),
\end{equation}
which applies to a single majoron mode with mass $m_J$. 
$n_J$ is the number density of the single majoron mode $J$ and $H$
is the Hubble parameter.
By summing over
all KK-modes of the majoron in the same way as it was done for gravitons
in Refs.~\cite{hs99,Hannestad:2001nq}, one finds a present day density of
\begin{eqnarray}
\rho_J & = & 8.3 \times 10^{-24} \frac{\pi^{\delta/2}}{\Gamma(\delta/2)}
\left(\frac{M_P}{T_{RH}}\right)^2
\left(\frac{T_{RH}}{M}\right)^{\delta+2} \,\, {\rm GeV}^4 \nonumber 
\vspace*{0.5cm} \\
&& \,\, \times \, g_{aa}^2 \,
\int_0^\infty dz z^{\delta-1} \int_z^\infty dq q^3 K_1(q)
\end{eqnarray}
Requiring that this density is smaller than the critical density,
$\rho_c = 8.1 \times 10^{-47} h^2 \,\, {\rm GeV}^4$, then yields an
upper bound on $g_{aa}$ as a function of $M$ and $T_{RH}$.

However, there is again a limit to the applicability of the bound
because for large $g_{aa}$ the majorons will have decayed by the
present. For the decay rate given in Eq.~(\ref{Eq:decay}) and the
requirement that $\tau \gtrsim 10^{10} \,\, {\rm yr}$ one finds
\begin{equation}
g_{aa} \lesssim 3 \times 10^{-19} T_{RH,{\rm MeV}}^{-1},
\end{equation}
assuming that the typical majoron mass is $\sim 3 T_{RH}$, a fairly
good approximation.
In Fig.~3 we show the contours of $g_{aa}$ corresponding to critical
density. Also shown is the lower bound on $M$ as a function of $T_{RH}$
from considering the decay of gravitons produced in the early
universe.
From this argument anything to the left of the thick grey lines
in the figure is excluded.
The limits used are the ones from Ref.~\cite{Hannestad:2001nq} which
are the most restrictive cosmological limits, combined with the
limits from Ref.~\cite{Hannestad:2001xi} which for low $T_{RH}$ can be
more restrictive. Finally, we also show the upper bound on
$g_{aa}$ from the above equation.
In the case where the line from the graviton bound is to the right of the
decay lifetime bound, majorons cannot possibly contribute to critical
density. This is seen to be the case for both $\delta=2$ and $\delta=6$,
and indeed for all values of $\delta$ from 2 to 6.
Therefore the conclusion is that no non-trivial bound on $g_{aa}$ comes
from cosmology, and further that majorons cannot contribute the dark
matter of the universe~\cite{valle}.
Already, gravitons are excluded as a dark matter candidate because
if they were to contribute critical density the photons produced by
their decay would have been clearly visible.

\begin{figure}[t]
\begin{center}
\includegraphics[width=7truecm]{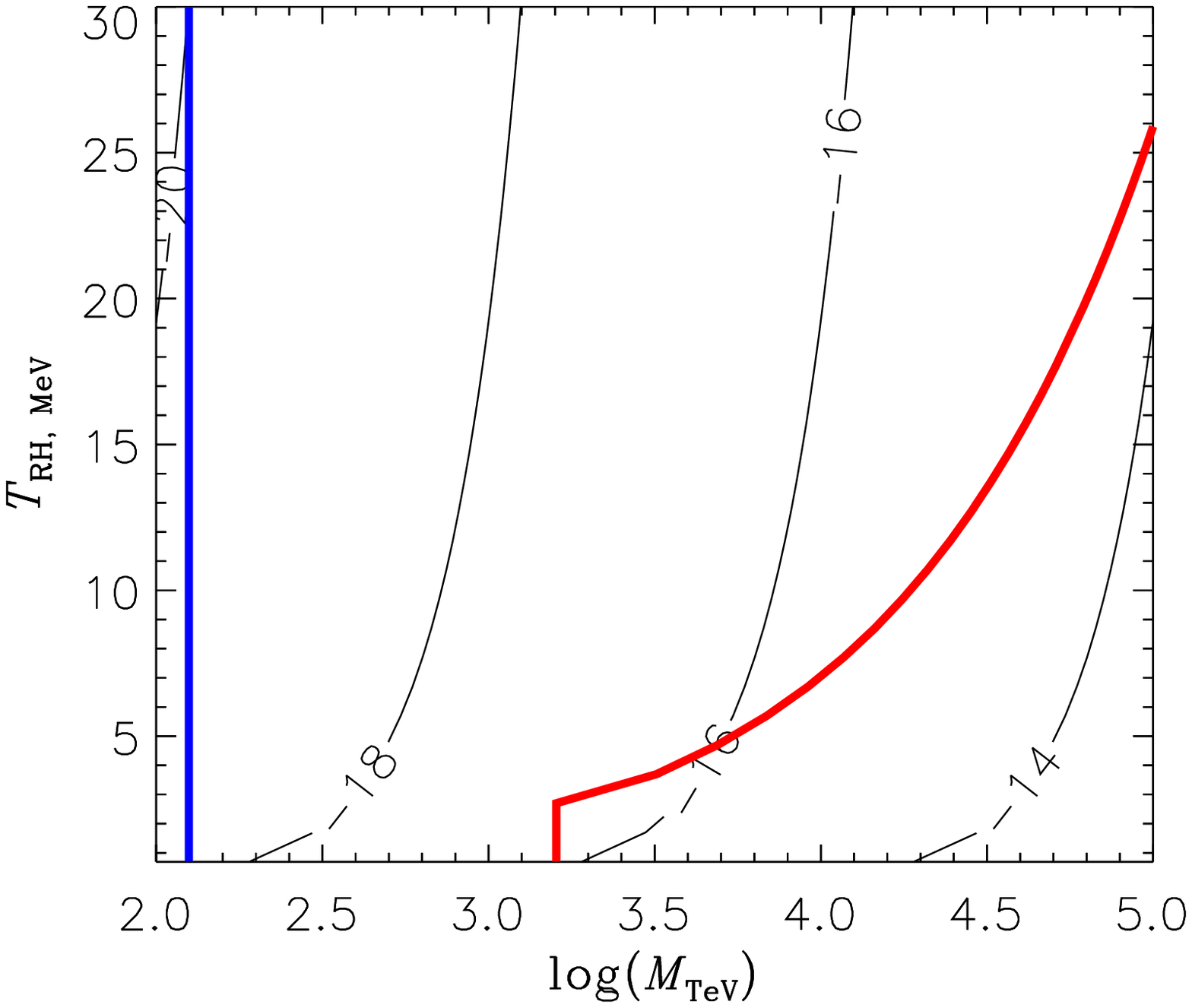}
\includegraphics[width=7truecm]{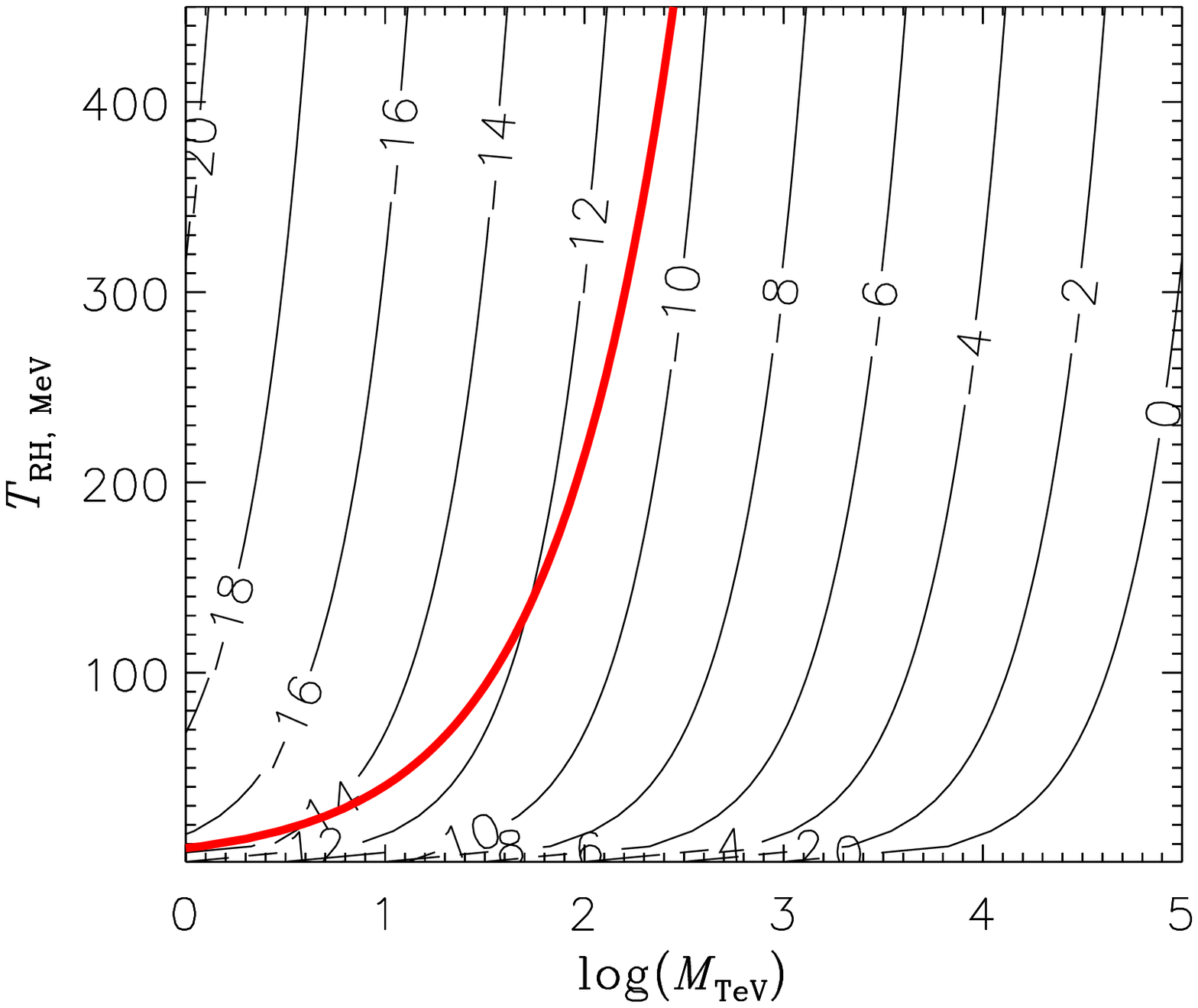}
\end{center}
\caption{Contours of $\log(g_{aa})$ which corresponds to $\rho_J = \rho_c$.
The upper panel is for $\delta=2$ and the lower for $\delta=6$.
The thick line to the right is the lower bound on $M$ coming from
considering graviton emission. The thick vertical line to
the left correponds to the
maximum $g_{aa}$ for which majorons live longer than the age of the universe.}
\label{fig4}
\end{figure}

\subsection{Conclusions}

In the present paper
we have shown that supernova constraints on brane-world
scenarios for neutrino masses are many orders of magnitude stronger
than the accelerator bound~\cite{MNSS}. 
Even so the constraints found are
somewhat weaker than what is naturally expected in see-saw
models of neutrino mass.

We have also shown that bulk majorons cannot act as the dark matter
in the universe, at least not within the present scenario with
equal radii of all the extra dimensions.

Finally, our findings suggest that
unnatural types of the ``see-saw" mechanism for neutrino masses are
unlikely to occur in nature, even in the presence of extra dimensions.

\acknowledgments
We wish to thank Jukka Maalampi, Georg Raffelt, Joseph Schechter, and
Jose Valle for valuable comments.
The work of F.S. is supported by the Marie--Curie
fellowship under contract MCFI-2001-00181.

\newpage


\begin{thebibliography}{99}
\bibitem{add98}N.~Arkani-Hamed, S.~Dimopoulos and G.~Dvali, Phys.\ Lett.\
{\bf B429}, 263 (1998).
\bibitem{Antoniadis}
I.~Antoniadis, N.~Arkani-Hamed, S.~Dimopoulos and G. Dvali,
Phys.\ Lett.\ B {\bf 436}, 257 (1998).
\bibitem{add99}N.~Arkani-Hamed, S.~Dimopoulos and G.~Dvali, Phys.\ Rev.\ D
{\bf 59}, 086004 (1999).

\bibitem{DDG} K. R. Dienes, E. Dudas and T. Gherghetta, Nucl. Phys.{\bf
B557},25(1999).

\bibitem{ADDM} N. Arkani-Hamed, S. Dimopoulos, G. Dvali and
J. March-Russell, hep-ph/9811448.

\bibitem{Mohapatra:1999zd}
R.~N.~Mohapatra, S.~Nandi and A.~Perez-Lorenzana,
Phys.\ Lett.\ B {\bf 466}, 115 (1999)
[arXiv:hep-ph/9907520].

\bibitem{Mohapatra:1999af}
R.~N.~Mohapatra and A.~Perez-Lorenzana,
Nucl.\ Phys.\ B {\bf 576}, 466 (2000)
[arXiv:hep-ph/9910474].

\bibitem{Dvali:1999cn}
G.~R.~Dvali and A.~Y.~Smirnov,
Nucl.\ Phys.\ B {\bf 563}, 63 (1999)
[arXiv:hep-ph/9904211].

\bibitem{Barbieri:2000mg}
R.~Barbieri, P.~Creminelli and A.~Strumia,
Nucl.\ Phys.\ B {\bf 585}, 28 (2000)
[arXiv:hep-ph/0002199].

\bibitem{Mohapatra:2000wn}
R.~N.~Mohapatra and A.~Perez-Lorenzana,
Nucl.\ Phys.\ B {\bf 593}, 451 (2001)
[arXiv:hep-ph/0006278].

\bibitem{Ioannisian:mu}
A.~Ioannisian and J.~W.~Valle,
Phys.\ Rev.\ D {\bf 63}, 073002 (2001).

\bibitem{FP} A. E. Faraggi and M. Pospelov, Phys. Lett. {\bf
B458},237(1999).

\bibitem{MN} G. C. McLaughlin and J. N. Ng, Phys. Lett. {\bf
B470},157(1999).

\bibitem{DK} A. Das and O. Kong, Phys. Lett. {\bf B470},149(1999).

\bibitem{C} C. D. Carone, Phys. Rev. {\bf D61},015008(2000).

\bibitem{BDN} T. Banks, M. Dine and A. E. Nelson, JHEP 9906: 014(1999).

\bibitem{Abazajian:2000hw}
K.~Abazajian, G.~M.~Fuller and M.~Patel,
arXiv:hep-ph/0011048.

\bibitem{Ioannisian:1999cw}
A.~Ioannisian and A.~Pilaftsis,
Phys.\ Rev.\ D {\bf 62}, 066001 (2000)
[arXiv:hep-ph/9907522].

\bibitem{MPP} R. N. Mohapatra, A. Perez-Lorenzano and C. A. de S. Pires,
Phys. Lett. {\bf B491},143(2000).

\bibitem{MNSS} S.~Moussa, S.~Nasri, F.~Sannino and J.~Schechter, hep-ph/0108128. To appear in Phys. Rev. D.

\bibitem{MRS} E. Ma, M. Raidal and U. Sarkar, Phys. Rev. Lett. {\bf
85},3769(2000).


\bibitem{rpp} Review of Particle Physics. D.E. Groom et al. Eur. Phys. J. {\bf C15},1(2000).

\bibitem{Carone:2002ss}
C.~D.~Carone, J.~M.~Conroy and H.~J.~Kwee,
Phys.\ Lett.\ B {\bf 538}, 115 (2002)
[arXiv:hep-ph/0204045].

\bibitem{TPV} R. Tomas, H. P{\"a}s and J. W. F. Valle, hep-ph/0103017.

\bibitem{CMP} Y. Chikashige, R. N. Mohapatra and R. D. Peccei, Phys. Lett.
{\bf B98},265(1981).

\bibitem{SV} J.~Schechter and J.W.F.~Valle, Phys.~Rev.~{\bf D25}, 774
(1982).

\bibitem{ss} T. Yanagida, Proc. of the Workshop on Unified Theory and
Baryon Number in the Universe, ed. by O. Sawada and A. Sugamato
(KEK Report 79-18,1979), p 95; M. Gell-Mann, P. Ramond and R.
Slansky in Supergravity, eds P. van Niewenhuizen and D. Z.
Freedman (North Holland, 1979);
 R. N. Mohapatra and G. Senjanovic,
Phys. Rev. Lett. \textbf{44}, 912 (1980).

\bibitem{Hirata:ad}
K.~S.~Hirata {\it et al.},
Phys.\ Rev.\ D {\bf 38} (1988) 448.

\bibitem{Bratton:ww}
C.~B.~Bratton {\it et al.}  [IMB Collaboration],
Phys.\ Rev.\ D {\bf 37} (1988) 3361.

\bibitem{baksan}E.~N.~Alexeyev {\it et al.},
Pis'ma Zh.\ Eksp.\ Teor.\ Fiz.\ {\bf 45} (1987) 461 [JETP Lett.\
{\bf 45} (1987) 589].


\bibitem{Aharonov:ee}
Y.~Aharonov, F.~T.~Avignone and S.~Nussinov,
Phys.\ Rev.\ D {\bf 37}, 1360 (1988).

\bibitem{Aharonov:ik}
Y.~Aharonov, F.~T.~Avignone and S.~Nussinov,
Phys.\ Rev.\ D {\bf 39}, 985 (1989).

\bibitem{Choi:1987sd}
K.~Choi, C.~W.~Kim, J.~Kim and W.~P.~Lam,
Phys.\ Rev.\ D {\bf 37}, 3225 (1988).

\bibitem{Grifols:1988fg}
J.~A.~Grifols, E.~Masso and S.~Peris,
Phys.\ Lett.\ B {\bf 215}, 593 (1988).

\bibitem{Berezhiani:gf}
Z.~G.~Berezhiani and M.~I.~Vysotsky,
Phys.\ Lett.\ B {\bf 199}, 281 (1987).

\bibitem{Berezhiani:za}
Z.~G.~Berezhiani and A.~Y.~Smirnov,
Phys.\ Lett.\ B {\bf 220}, 279 (1989).

\bibitem{Choi:1989hi}
K.~Choi and A.~Santamaria,
Phys.\ Rev.\ D {\bf 42}, 293 (1990).

\bibitem{Chang:1993yp}
S.~Chang and K.~Choi,
Phys.\ Rev.\ D {\bf 49}, 12 (1994)
[arXiv:hep-ph/9303243].

\bibitem{Pilaftsis:1993af}
A.~Pilaftsis,
Phys.\ Rev.\ D {\bf 49}, 2398 (1994)
[arXiv:hep-ph/9308258].


\bibitem{Berezhiani:1994jc}
Z.~G.~Berezhiani and A.~Rossi,
Phys.\ Lett.\ B {\bf 336}, 439 (1994)
[arXiv:hep-ph/9407265].

\bibitem{Kachelriess:2000qc}
M.~Kachelriess, R.~Tomas and J.~W.~Valle,
Phys.\ Rev.\ D {\bf 62}, 023004 (2000)
[arXiv:hep-ph/0001039].

\bibitem{Tomas:2001dh}
R.~Tomas, H.~P{\"a}s and J.~W.~Valle,
Phys.\ Rev.\ D {\bf 64}, 095005 (2001)
[arXiv:hep-ph/0103017].



\bibitem{Raffelt:wa}
G.~G.~Raffelt,
``Stars As Laboratories For Fundamental Physics: The Astrophysics Of Neutrinos, Axions, And Other Weakly Interacting Particles,''
{\it  Chicago, USA: Univ. Pr. (1996) 664 p}.


\bibitem{Cullen:1999hc}
S.~Cullen and M.~Perelstein,
Phys.\ Rev.\ Lett.\  {\bf 83}, 268 (1999)
[arXiv:hep-ph/9903422].

\bibitem{Barger:1999jf}
V.~D.~Barger, T.~Han, C.~Kao and R.~J.~Zhang,
Phys.\ Lett.\ B {\bf 461}, 34 (1999)
[arXiv:hep-ph/9905474].

\bibitem{hanhart}
C.~Hanhart, D.~R.~Phillips, S.~Reddy and M.~J.~Savage,
Nucl.\ Phys.\ B {\bf 595}, 335 (2001)
[arXiv:nucl-th/0007016].

\bibitem{hanhart2}
C.~Hanhart, J.~A.~Pons, D.~R.~Phillips and S.~Reddy,
Phys.\ Lett.\ B {\bf 509}, 1 (2001)
[arXiv:astro-ph/0102063].


\bibitem{Hannestad:2001jv}
S.~Hannestad and G.~Raffelt,
Phys.\ Rev.\ Lett.\  {\bf 87}, 051301 (2001)
[arXiv:hep-ph/0103201].

\bibitem{Hannestad:2001xi}
S.~Hannestad and G.~G.~Raffelt,
Phys.\ Rev.\ Lett.\  {\bf 88}, 071301 (2002)
[arXiv:hep-ph/0110067].

\bibitem{egret}D.~A.~Kniffen {\it et al.},
Astron.\ Astrophys.\ Suppl.\ {\bf 120}, 615 (1996).

\bibitem{0953}J.~D.~Pilkington {\it et al.},
Nature {\bf 218}, 126 (1968).

\bibitem{psc96}G.~G.~Pavlov, G.~S.~Stringfellow and F.~A.~Cordova,
Astrophys.\ J.\ {\bf 467}, 370 (1996).

\bibitem{ll99}M.~B.~Larson and B.~Link,
Astrophys.\ J.\ {\bf 521}, 271 (1999).

\bibitem{hr02}The same problem exists for gravitons. However, an
actual calculation for gravitons again turns out to give a
reabsorption timescale much longer than the age of PSR 0953+755
(S.~Hannestad and G.~Raffelt, in preparation).

\bibitem{hs99}
L.~J.~Hall and D.~R.~Smith,
Phys.\ Rev.\ D {\bf 60}, 085008 (1999)
[arXiv:hep-ph/9904267].

\bibitem{Hannestad:2001nq}
S.~Hannestad,
Phys.\ Rev.\ D {\bf 64}, 023515 (2001)
[arXiv:hep-ph/0102290].

\bibitem{Fairbairn:2001ct}
M.~Fairbairn,
Phys.\ Lett.\ B {\bf 508}, 335 (2001)
[arXiv:hep-ph/0101131].

\bibitem{Fairbairn:2001ab}
M.~Fairbairn and L.~M.~Griffiths,
JHEP {\bf 0202}, 024 (2002)
[arXiv:hep-ph/0111435].

\bibitem{Kawasaki:1999na}
M.~Kawasaki, K.~Kohri and N.~Sugiyama,
Phys.\ Rev.\ Lett.\  {\bf 82}, 4168 (1999)
[arXiv:astro-ph/9811437].


\bibitem{Kawasaki:2000en}
M.~Kawasaki, K.~Kohri and N.~Sugiyama,
Phys.\ Rev.\ D {\bf 62}, 023506 (2000)
[arXiv:astro-ph/0002127].

\bibitem{valle}It should be noted that 3+1 dimensional majoron models might
still be of relevance for the dark matter problem.
V.~Berezinsky and J.~W.~F.~Valle, Phys.\ Lett.\ {\bf B318}, 360 (1993).



\end{thebibliography}
\end{document}